\documentclass[fleqn,usenatbib,useAMS]{mnras}

\usepackage{mathptmx}
\usepackage{txfonts}

\usepackage[T1]{fontenc}
\usepackage{ae,aecompl}

\DeclareRobustCommand{\VAN}[3]{#2}
\let\VANthebibliography\thebibliography
\def\thebibliography{\DeclareRobustCommand{\VAN}[3]{##3}\VANthebibliography}

\usepackage{graphicx}	
\usepackage{bm}
\usepackage{widetext}
\usepackage{subfig,epsfig,epsf,color}
\usepackage[utf8x]{inputenc}
\usepackage{appendix}
\usepackage{slashed}
\usepackage{feynmf}
\usepackage{array}
\usepackage{pdflscape}	
\usepackage{hyperref}

\newcommand{\bea}{\begin{eqnarray}}
\newcommand{\eea}{\end{eqnarray}}

\title[Implications of Interacting DM on NS]{Implications of Feebly Interacting Dark Sector on Neutron Star Properties and Constraints from GW170817}

\author[Debashree Sen and Atanu Guha]{
Debashree Sen,$^{1,2}$\thanks{E-mail: debashree@iiserbpr.ac.in}
Atanu Guha,$^{3}$\thanks{E-mail: atanu.guha@students.iiserpune.ac.in}
\\
$^{1}$Department of Physical Sciences\\ Indian Institute of Science Education and Research Berhampur,\\
Transit Campus, Government ITI, Berhampur-760010, Odisha, India\\
$^{2}$Physics Group, Variable Energy Cyclotron Centre, 1/AF Bidhan Nagar, Kolkata
700064, India\\
$^{3}$Department of Physical Sciences\\
Indian Institute of Science Education and Research Pune,\\
Dr. Homi Bhabha Road, Ward No. 8, NCL Colony, Pashan, Pune-411008, Maharashtra, India
}

\date{Accepted XXX. Received YYY; in original form ZZZ}

\pubyear{2021}

\begin{document}
\label{firstpage}
\pagerange{\pageref{firstpage}--\pageref{lastpage}}
\maketitle

\begin{abstract}

We investigate the effect of feeble interaction of dark matter (DM) with hadronic matter on the equation of state (EoS) and structural properties of neutron stars (NSs) in static conditions. For the purpose we adopt the effective chiral model for the hadronic sector and for the first time in the context of possible existence of DM inside NSs, we introduce DM-SM interaction through light new physics mediator. Moreover, the mass of DM fermion, the mediator and the coupling are adopted from the self-interaction constraint from Bullet cluster and from present day relic abundance. Within the considered framework, the work highlights the underlying stiffening of EoS in presence of DM fermion of mass of the order of a few GeV compared to the no-DM scenario. Consequently, the maximum gravitational mass of NS is obtained consistent with the bounds from the most massive pulsars which were not satisfied with the hadronic matter EoS alone. The estimates of radius and tidal deformability of 1.4 $M_{\odot}$ NS and the tidal deformabilities of the individual components of the binary neutron stars (BNS) associated with GW170817 are all in good agreement with the individual constraints obtained from GW170817 observation of BNS merger.


\end{abstract}




\section{Introduction}
\label{intro}

 Robust evidences like rotation curves of galaxies, observation of gravitational lensing, x-ray analysis of Bullet cluster affirm the existence of dark matter (DM) in our Universe~\cite{Bertone:2004pz, Aghanim:2018eyx}. The parameters of the standard model of cosmology are measured using the analysis of the Cosmic Microwave Background (CMB) anisotropy maps. CMB maps are obtained from the Wilkinson Microwave Anisotropy Probe (WMAP) data \cite{Ade:2013zuv, Bennett:2012zja}. These measurements and subsequent analysis suggest that the contribution of DM to the total matter content of the Universe to be $\sim 26\%$, whereas, only $\sim 4\%$ of the Universe is constituted by the baryonic matter. The understanding of standard cosmology furnishes the present day relic abundances of DM to be $\Omega h^2 \approx 0.12$ with an uncertainty at the level of $1\%$ \cite{Cannoni:2015wba}. N-body simulations show that the DM profile is Universal i.e., same for all masses~\cite{Navarro:1995iw}. 
 
 Most popular particle candidates for DM are characterized by their feeble interaction with the standard model (SM) particles and by their massiveness to generate high gravitational force. The search for such weakly interacting massive particles (WIMPs) are going on over a few decades, specifically following certain avenues. One of them is the direct detection, in which DM particles scatter of nuclei and nuclear/electron recoil energies are measured. Dedicated experiments for direct detection of DM candidates are DAMA/LIBRA \cite{Bernabei:2008yi,Aalseth:2008rx,Aalseth:extra}, KIMS \cite{Lee:2014zsa}, CRESST-II \cite{Angloher:2015ewa}, ZEPLIN \cite{Sumner:2005wu}, CRESST-I \cite{Bravin:1999fc}, superCDMS \cite{Agnese:2018col}, XENON100 \cite{Aprile:2012nq}, XENON1T \cite{Aprile:2018dbl}, LUX \cite{Akerib:2012ak}, PANDAX-II \cite{Wang:2020coa}, DARKSIDE-50 \cite{Agnes:2018oej}, SENSEI \cite{Crisler:2018gci}, ADMX \cite{Du:2018uak}. Very recently in 2020 the XENON1T observed excess over predicted signal \cite{Aprile:2020tmw}. The XENON1T collaboration reported an $3.5\sigma$ excess of events in the electron recoil range of $1~\rm{keV} < E_R < 7~\rm{keV}$ with 285 events over the backgrounds of $232 \pm 15$ events, with an exposure of 0.65 tonne-years and an unprecedentedly low background \cite{Aprile:2020tmw}. While the other bins were nearly consistent with the expected background events, the prominent excess events appeared in the 2-3 keV bins. Another avenue is the indirect detection, where DM particles annihilate via self interaction and SM remnants are looked for.  For indirect searches, the leading experiments are FERMI-LAT \cite{Atwood:2009ez}, DAMPE \cite{Ambrosi:2017wek}, IceCube \cite{Aartsen:2014gkd}, ATIC \cite{Chang:2008aa}, PAMELA \cite{Adriani:2008zr, Adriani:2013uda}, AMS-02 \cite{Accardo:2014lma, Aguilar:2014mma}, Voyager1 \cite{Boudaud:2016mos}, CALET \cite{Adriani:2017efm, Adriani:2018ktz}, HAWC \cite{Harding:2015bua}, HESS \cite{Abdallah:2016ygi}, VERITAS \cite{Zitzer:2017xlo}, MAGIC \cite{Elsaesser:2004xa}, CTA (proposed) \cite{Carr:2015hta}.
Other avenues include collider searches (where, excess of predicted signals are analyzed to interpret the annihilation of SM particles to produce DM via new physics channel) \cite{Aaboud:2019opc,Aaboud:2018jbr,Aaboud:2018aqj,Aaboud:2018arf,Aaboud:2018hdl,Aaboud:2018iil,Aaboud:2017iio,Aad:2019tua,Aad:2019xav,Aad:2020srt,Aad:2019tcc,Aad:2015rba,Sirunyan:2019xwh,Sirunyan:2020pjd,Sirunyan:2018vlw,Sirunyan:2018ryt,Sirunyan:2019wau,Sirunyan:2019gut,Sirunyan:2018ldc,Sirunyan:2018pwn,Khachatryan:2016sfv,Alimena:2019zri,Bhattacherjee:2020nno,Aaij:2020iew,Blekman:2020hwr}, present day non-baryonic relic abundances \cite{Ade:2013zuv, Bennett:2012zja} and CMB spectral distortion data by FIRAS and PIXIE \cite{Ali-Haimoud:2015pwa}.

 Due to having weak interactions with SM particles, WIMPs were in thermal equilibrium with the plasma in early Universe. Assuming a dark matter candidate with mass much less than electroweak scale, one can find that the annihilation of the WIMPs before freeze-out was mediated by weak interactions only and the theoretically predicted relic abundances are in good agreement with the observation of total non-baryonic relic density ($\approx 0.12 $) of the Universe at present day \cite{Plehn:2017fdg}, which has been obtained from the measurement of CMB anisotropy and the spatial distribution of galaxies \cite{Zyla:2020zbs, Tanabashi:2018oca}. In DM phenomenology, this outcome is usually referred to as the WIMP miracle. Any DM model must reproduce this relic density. This requirement sets strong constraints on the model parameters. 
 
 In literature there is a wide mass range for DM particle candidates, it is spread from $10^{-22}~\rm{eV}$ (fuzzy DM)\cite{Ni:2019qfa,Bauer:2020zsj,Davoudiasl:2017jke,Bernal:2017oih} to $10^{15}~\rm{GeV}$ \cite{Kolb:2007vd,Kolb:2017jvz,Alcantara:2019sco}. Pauli exclusion principle forbids the mass of fermionic DM to be less than $25~\rm{eV}$ \cite{Baltz:2004tj, Tremaine:1979we}. However, ideally the candidates of mass below few keV are too hot for structure formation while for those above $100~\rm{TeV}$, the perturbative unitarity violates.

 On the other hand, neutron stars (NSs) can be treated as natural laboratories where several branches of physics can be explored. Such branches include the study of matter at extreme conditions of density (5-10 times saturation density) and even beyond the standard practice, be it standard cosmology, gravity or the standard model (SM) of particle physics. The equation of state (EoS) at such high density depends on the composition of NS matter (NSM) which is largely unknown from experimental perspectives. However, few observational and empirical constraints on the structural properties of NS help us to constraint the EoS to certain extent. Massive pulsars like PSR J0348+0432 \cite{Antoniadis:2013pzd} and PSR J0740+6620 \cite{Cromartie:2019kug} have put strong upper bounds on the gravitational mass of NSs. With the detection of gravitational wave (GW170817) from binary NS merger (BNSM), stringent constraints on the radius and tidal deformability of a 1.4 $M_{\odot}$ NS are also obtained. Moreover, NICER experiment came up with constraints on mass-radius relationship for PSR J0030+0451 \cite{Riley:2019yda,Miller:2019cac}. Also from the source spectrum analysis of 1E 1207.4-5209 \cite{Sanwal:2002jr} and RX J0720.4-3125 \cite{Hambaryan:2017wvm}, the maximum bounds on surface redshift are obtained. Any realistic EoS obtained by theoretical modeling of NSM must satisfy the aforesaid constraints on the structural properties of NSs.
 
 NSs are remnants of massive stars that have gone through various stages after the exhaustion of thermo-nuclear fuel. One such prominent stage is the supernova (SN) explosion. Hypothesized DM-SM particles interaction indicates that feasible amount of non-thermal production of DM may take place during the SN explosion. Energy released in a SN explosion is carried away by neutrinos whereas the careful analysis of Kamiokande~\cite{Hirata:1987hu} and IMB(Irvine-Michigan-Brookhaven)~\cite{Bionta:1987qt} data on SN1987A suggests the possible existence of some new physics channel contributing to the SN cooling scenario. But these beyond standard model (BSM) channels are highly constrained~\cite{Raffelt:1996wa}. Sub-GeV DM particles can contribute to the SN cooling scenario but DM particles of mass more than $100~\rm{MeV}$ are phase-space suppressed~\cite{Kadota:2014mea, Dreiner:2013mua, Guha:2018mli, Guha:2015kka}. If we consider GeV scale DM particles to take part in SN cooling, our current understanding of the cooling mechanism and neutrino signal from SN1987A would be invalidated~\cite{Janka:2006fh, Janka:2017vlw}. Also the environment could not afford to lose much energy on production of such DM at that stage in order to drive the successful SN explosion. Sub-GeV DM can also get trapped depending on the coupling strength if they do not satisfy the optical depth criteria of free streaming~\cite{Guha:2018mli, Shapiro:1983du}. But the amount of such DM trapping can safely be assumed to be small due to much lower production crossection  as supported by the direct and indirect detection experiments~\cite{Essig:2012yx, Crisler:2018gci, Agnese:2018col, Boudaud:2016mos, Ali-Haimoud:2015pwa}. Hence in this work we will consider only thermal DM as a possible constituent for the NSM~\cite{Panotopoulos:2017idn, Li:2012ii, Bertoni:2013bsa}.
 
 NSs can inherit the trapped DM during SN explosion or can accrete \cite{Razeira:2011zza,PerezGarcia:2010ap,deLavallaz:2010wp,Ciarcelluti:2010ji}. They can also create their own DM but the amount is negligible after the Kelvin-Helmholtz cooling phase~\cite{Bertoni:2013bsa, Nelson:2018xtr}. In our present work we considered that the accretion of thermal DM during the NS formation is dominant compared to the other possible sources. Theoretically, the presence of DM in NSs can be explained either by considering the interaction of the DM particles with the baryons (via exchange of Higgs boson) \cite{Panotopoulos:2017idn, Bertoni:2013bsa, Nelson:2018xtr} or without considering any particle interaction between DM and baryonic matter \cite{Ellis:2018bkr, Li:2012ii}. For the former type of treatment, the DM-bayonic matter interaction is implemented in the Lagrangian of the total NS matter including DM and baryonic matter while for the latter type of treatment, the two types of matter interact only gravitationally and the two-fluid method is often successfully adopted \cite{Tolos:2015qra, Deliyergiyev:2019vti, Rezaei:2016zje, Mukhopadhyay:2016dsg}. It is seen that MeV-GeV scale DM particles may be capable of playing a key role in BNSM. The dimensionless tidal deformability ($\Lambda$) is enhanced significantly by the presence of trace amount of DM interacting with baryonic matter inside NSs \cite{Bertoni:2013bsa, Nelson:2018xtr}. However, in the case where DM does not interact with hadronic matter, the tidal deformability is seen to decrease considerably \cite{Ellis:2018bkr} along with mass and radius \cite{Li:2012ii}.  In the present work, we consider the former type of treatment where fermionic DM particles $\chi$ interact feebly with baryonic matter $\psi$ via exchange of a light and scalar new physics mediator $\phi$ that mixes with SM Higgs boson. The main aim of the present work is to study the effects of interaction of DM and baryonic matter on NS properties of DM admixed NSs.  
 
 In popular models in literature, where DM-baryonic matter interaction is taken into account, DM directly interacts with nucleons via the exchange of SM Higgs Boson, irrespective of its spin~\cite{Andreas:2008xy}. Fundamentally there is no theoretical justification for that and in principle DM-Higgs coupling can be zero~\cite{Panotopoulos:2017idn}. In the following discussion ahead, we considered a singlet scalar mediator belonging to new physics, which mixes with SM-Higgs boson and hence be able to interact to the nucleons~\cite{Krnjaic:2015mbs}. Presence of light GeV-scale DM without interactions with the nucleons can actually soften the EoS drastically~\cite{Li:2012ii}, whereas by introducing the interaction term we obtained stiffer EoS (Fig. \ref{eos1}). For the pure hadronic matter sector, we have adopted the effective chiral model \cite{Sahu:2000ut,Jha:2009kt,Sen:2018yyq,Sen:2020edi} for $\beta$ equilibrated NSM. However, for a two-fluid treatment, where interaction between DM and HM is not considered via particle interaction, further realistic EoS for the hadronic matter may also be adopted. For example, EoS obtained by interpolating high and low density domains and constrained by the inputs from low-energy nuclear physics, high-density limit from perturbative QCD, and observational data of pulsars \cite{Kurkela:2014vha}, is successfully adopted to describe NS configurations in the presence of DM with the help of two-fluid approach \cite{Tolos:2015qra, Deliyergiyev:2019vti}. 
 
 Our discussion is organized as follows. In Sec.~\ref{formalism} we present the basic details of the model in presence of DM and the parameter sets we considered. Numerical results and subsequent discussions can be found in Sec.~\ref{Results}. Finally we conclude in Sec.~\ref{Conclusions}.

\section{Formalism}
\label{formalism}

\subsection{Model including Higgs portal dark sector}

 We consider the effective chiral model \cite{Sahu:2000ut,Jha:2009kt,Sen:2020edi} for the hadronic matter sector. The model is based on chiral symmetry and the interaction of the nucleons $\psi$ with the scalar $\sigma$, vector $\omega$, isovector $\rho$ and the pseudoscalar $\pi$ mesons as mediators. In addition to the Lagrangian density for the effective chiral model given in~\cite{Sahu:2000ut,Jha:2009kt,Sen:2020edi}, we propose to include the following renormalizable operator to incorporate the new physics (feebly interacting dark sector) 
 \bea
 \mathcal{L}_{new} = \partial_\mu \Phi \partial^\mu \Phi - \mu_\Phi^2 \Phi^\dagger \Phi - \lambda_\Phi \left(\Phi^\dagger \Phi\right)^2 - \lambda_{mix} \Phi^\dagger \Phi H^\dagger H 
 \label{Eq:lnew}
 \eea
 
 where, $\Phi$ is the singlet scalar mediator of mass $m_\phi$. For this theory we can diagonalize the scalar mass terms (Eq.(\ref{Eq:lnew})) after electroweak symmetry breaking. The resulting mass eigenstates are identified as follows: $\phi$ be the DM-SM mediator
and $h$ the Higgs boson~\cite{Krnjaic:2015mbs}. Invoking this, total Lagrangian density becomes
\bea
\mathcal{L} = \mathcal{L}_{hadron} + \mathcal{L}_{dm}
\eea
where,
\bea
\mathcal{L}_{dm} = \frac{1}{2} \partial_\mu \phi \partial^\mu \phi - \frac{1}{2}m_\phi^2 \phi^2 - \frac{1}{4}\lambda_\phi \phi^4 +  \bar{\chi} \left[i \gamma_\mu \partial^\mu - m_\chi \right] \chi - y_\phi \phi \bar{\chi} \chi 
\eea

 where, $\chi$ is the DM fermion of mass $m_\chi$ and it couples to the scalar mediator $\phi$ via Yukawa type coupling with coupling strength $y_{\phi}$. Also the new mediator acquires a coupling to SM fermions as it mixes with the Higgs boson. Expanding in mass basis, we obtain the mediator-SM interaction 
 \bea
 \mathcal{L}_{\phi, SM} = \sum_f g_f \bar{f} f \phi
 \eea 
 
 where, $f$ be the SM fermions of mass $m_f$ and the corresponding coupling
 \bea
 g_f = \frac{m_f}{v} \sin \theta
 \eea 
 
 $\theta$ be the mixing parameter and $v = 246~\rm{GeV}$ be the vacuum expectation value of the Higgs boson. From the coupling of the $\phi$ to the quarks we can estimate the effective coupling of the $\phi$ to the nucleons, let us call it $g_{\phi N N}$ (details given in the appendix).
 
 Now taking the contribution of the SM and BSM particles, the total Lagrangian density becomes
 
 \begin{widetext}
 \bea
 \mathcal{L} &=& \bar{\psi} \left[ \left( i \gamma_\mu \partial^\mu - g_\omega \gamma_\mu \omega^\mu - \frac{1}{2} g_\rho \vec{\rho_\mu}\cdot \vec{\tau} \gamma^\mu \right) - g_\sigma \left( \sigma + i \gamma_5 \vec{\tau}\cdot \vec{\pi} \right) -g_{\phi N N} \phi \right] \psi \nonumber \\
 &+& \frac{1}{2} \left(\partial_\mu \vec{\pi} \partial^\mu \vec{\pi} + \partial_\mu \sigma \partial^\mu \sigma \right) - \frac{\lambda}{4} \left(x^2 - x_0^2 \right)^2 - \frac{\lambda B}{6} \left(x^2 - x_0^2 \right)^3 - \frac{\lambda C}{8} \left(x^2 - x_0^2 \right)^4 - \frac{1}{4} F_{\mu \nu} F^{\mu \nu} + \frac{1}{2} g_\omega^2 x^2 \omega_\mu \omega^\mu - \frac{1}{4} \vec{R}_{\mu \nu} \cdot \vec{R}^{\mu \nu} + \frac{1}{2} m_\rho^2 \vec{\rho_\mu}\cdot \vec{\rho^\mu} \nonumber \\
 &+& \frac{1}{2} \partial_\mu \phi \partial^\mu \phi - \frac{1}{2}m_\phi^2 \phi^2 - \frac{1}{4}\lambda_\phi \phi^4 +  \bar{\chi} \left[i \gamma_\mu \partial^\mu - \left(m_\chi + y_\phi \phi \right) \right] \chi 
 \label{Eq:Lagrangian}
 \eea
  \end{widetext}

We have the following relations from chiral symmetry

\bea
m &=& g_{\sigma} x_0 + g_{\phi NN} \phi_0; ~~~ m_{\sigma} = \sqrt{2 \lambda}~ x_0; ~~~ m_{\omega}=g_{\omega} x_0 \nonumber \\
\left\langle \sigma \right\rangle &=& x_0;~~~ C_{\omega} = \frac{g^2_{\omega}}{m^2_{\omega}};~~~ C_{\sigma} = \frac{g^2_{\sigma}}{m^2_{\sigma}}
\eea

$B$ and $C$ are coefficients of the higher order scalar field terms. Like \cite{Panotopoulos:2017idn} we have taken attractive potential for the new physics scalar mediator consistent with \cite{Hambye:2019tjt,Barbieri:1988ct}. Note that in the total Lagrangian, the last term $\bar{\psi}(g_{\phi N N} \phi)\psi$ of the first line of equation \ref{Eq:Lagrangian} indicates the interaction between the scalar new physics mediator $\phi$ from dark sector and the nucleons $\psi$. The strength of the interaction is determined by the coupling constant $g_{\phi N N}$ which is evaluated in the appendix.

 The effective masses of nucleons ($m^\star$) and dark matter ($m^\star_\chi$) are obtained as
\bea
m^\star &=& g_\sigma \sigma + g_{\phi N N} \phi \nonumber \\
m^\star_\chi &=& m_\chi + y_\phi \phi
\eea

 The equation of motion for the different mesons, nucleons and the DM can be obtained by applying the mean field treatment. The scalar equation of motion is given in terms of $Y=m^{\star}/m$ as

\begin{widetext} 
\bea
(1 - Y^2) -\frac{B}{C_{\omega}}(1-Y^2)^2 +\frac{C}{C_{\omega}^2}(1-Y^2)^3 -\frac{2C_{\sigma}\rho_s}{Y(m - g_{\phi NN}\phi_0)} +\frac{2C_{\sigma}C_{\omega}}{Y^4(m - g_{\phi NN}\phi_0)^2} = 0
\label{scalar_field}
\eea
\end{widetext}

while the vector $\omega$ and isovector $\rho$ meson field equations remain unchanged and same as obtained without including DM \cite{Sahu:2000ut,Jha:2009kt} viz.

\bea
\omega_0=\frac{\rho}{g_{\omega} x^2}
\label{vector_field}
\eea

and
\bea
\rho_{03}=\sum_{N}\frac{g_{\rho}}{m_\rho^2}I_{3_N}\rho_N 
\label{isovector_field}
\eea

Here `3' denotes for the third component in isospin $I$ of the individual nucleons $N$. 

 The scalar density is given as
\bea
\rho_S=\left\langle\overline{\psi}\psi\right\rangle=\frac{\gamma}{2 \pi^2} \int^{k_F}_0 dk ~k^2 \frac{m^{\star}}{\sqrt{k^2 + {m^{\star}}^2}}
\eea

while the baryon density as

\bea
\rho=\left\langle\psi^\dagger\psi\right\rangle=\frac{\gamma}{2\pi^2} \int^{k_F}_0 dk ~k^2  
\eea

Here the total baryon density is the sum of individual nucleon densities i.e., $\rho=\rho_n + \rho_p$.

 Since in the mean field treatment we have $\left\langle\pi\right\rangle=0$ with $m_{\pi}=0$, the explicit contribution of the pions is not taken into account. 

 The equation of motion of the DM-SM mediator field is given as

\bea
\phi_0=\frac{m^\star_\chi-m_\chi}{y_{\phi}}
\eea

\subsection{Equation of State}
 
 The EoS viz. the energy density $\varepsilon$ and pressure $P$ is calculated by calculating the energy-momentum tensor involving the Lagrangian \ref{Eq:Lagrangian}. The total energy density and total pressure is computed as

\begin{widetext}
\bea
\varepsilon &=& \frac{\left(m - g_{\phi NN}\phi_0\right)^2 \left(1-Y^2\right)^2}{8 C_{\sigma}} - \frac{\left(m - g_{\phi NN}\phi_0\right)^2 B \left(1-Y^2\right)^3}{12 C_{\sigma} C_{\omega}} + \frac{\left(m - g_{\phi NN}\phi_0\right)^2 C \left(1-Y^2\right)^4}{16 C_{\sigma} C_{\omega}^2} + \frac{C_{\omega} \rho^2}{2 Y^2}  \nonumber \\
&+&  \frac{1}{2} m_\rho^2 \rho_{03}^2 + \frac{\gamma}{2 \pi^2} \int_0^{k_F} \sqrt{k^2 + {m^{\star}}^2}~ k^2 dk + \frac{\gamma}{2 \pi^2} \sum_{\lambda^\prime=e,\mu} \int_0^{k_{\lambda^\prime}}  \sqrt{k^2 + m_{\lambda\prime}^2}~ k^2 dk  + \frac{1}{2} m_\phi^2 \phi_0^2 + \frac{1}{4} \lambda_\phi \phi_0^4 + \frac{\gamma_\chi}{2 \pi^2} \int_0^{k_F^{\chi}} \sqrt{k_\chi^2 + {m^{\star}}_\chi^2}~ k_\chi^2 dk_\chi
\label{e}
\eea
\\
\bea
P &=& - \frac{\left(m - g_{\phi NN}\phi_0\right)^2 \left(1-Y^2\right)^2}{8 C_{\sigma}} + \frac{\left(m - g_{\phi NN}\phi_0\right)^2 B \left(1-Y^2\right)^3}{12 C_{\sigma} C_{\omega}} - \frac{\left(m - g_{\phi NN}\phi_0\right)^2 C \left(1-Y^2\right)^4}{16  C_{\sigma} C_{\omega}^2} + \frac{C_{\omega} \rho^2}{2 Y^2} \nonumber \\
&+&  \frac{1}{2} m_\rho^2 \rho_{03}^2 + \frac{\gamma}{6 \pi^2} \int_0^{k_F} \frac{k^4 dk}{\sqrt{k^2 + {m^{\star}}^2}} + \frac{\gamma}{6 \pi^2} \sum_{\lambda^\prime=e,\mu} \int_0^{k_{\lambda^\prime}}  \frac{k^4 dk}{\sqrt{k^2 + m_{\lambda^\prime}^2}}  - \frac{1}{2} m_\phi^2 \phi_0^2 - \frac{1}{4} \lambda_\phi \phi_0^4 + \frac{\gamma_\chi}{6 \pi^2} \int_0^{k_F^{\chi}} \frac{k_\chi^4 dk_\chi}{\sqrt{k_\chi^2 + {m^{\star}}_\chi^2}}
\label{P}
\eea 
\end{widetext}

\subsection{Model parameter set for hadronic matter}
\label{Param_HM}

There are five parameters of the hadronic model given as $C_i=g_i^2/m_i^2$ where $i=\sigma,\omega,\rho$ and $B$ and $C$. They are determined by reproducing the saturated nuclear matter (SNM) properties. The hadronic model parameter set, chosen for the present work, is adopted from \cite{Jha:2009kt} and is given in table \ref{table_NM} along with the SNM properties yielded by the set.

\begin{table*}
\caption{Parameter set for the hadronic model (adopted from \protect\cite{Jha:2009kt}) along the saturation properties.}
{{
\setlength{\tabcolsep}{3.5pt}
\centering
\begin{tabular}{cccccccccccccc}
\hline
\hline
\multicolumn{1}{c}{$C_{\sigma}$}&
\multicolumn{1}{c}{$C_{\omega}$} &
\multicolumn{1}{c}{$C_{\rho}$} &
\multicolumn{1}{c}{$B/m^2$} &
\multicolumn{1}{c}{$C/m^4$} &
\multicolumn{1}{c}{$Y$} &
\multicolumn{1}{c}{$m_{\sigma}$} &
\multicolumn{1}{c}{$f_{\pi}$} &
\multicolumn{1}{c}{$K$} & 
\multicolumn{1}{c}{$B/A$} &
\multicolumn{1}{c}{$J(L_0)$} &
\multicolumn{1}{c}{$\rho_0$} \\
%
\multicolumn{1}{c}{($fm^2$)} &
\multicolumn{1}{c}{($fm^2$)} &
\multicolumn{1}{c}{($fm^2$)} &
\multicolumn{1}{c}{($fm^2$)} &
\multicolumn{1}{c}{($fm^4$)}&
\multicolumn{1}{c}{} &
\multicolumn{1}{c}{(MeV)} &
\multicolumn{1}{c}{(MeV)} &
\multicolumn{1}{c}{(MeV)} &
\multicolumn{1}{c}{(MeV)} &
\multicolumn{1}{c}{(MeV)} &
\multicolumn{1}{c}{($fm^{-3}$)} \\
\hline
\hline

7.325  &1.642  &5.324 &-6.586   &0.571    &0.87  &444.614   &153.984  &231  &-16.3   &32(88)  &0.153 \\ 
\hline
\hline
\end{tabular}
}}
\protect\label{table_NM}
\end{table*}

 The values of SNM properties like symmetry energy coefficient ($J = 32$~MeV), saturation density ($\rho_0 = 0.153$~$\rm{fm^{-3}}$), binding energy per particle ($B/A = -16.3$~MeV) and the nuclear incompressibility ($K = 231$~ MeV), yielded by the model parameter are well consistent with findings of \cite{Stone:2006fn,Dutra:2014qga,Khan:2012ps,Khan:2013mga,Garg:2018uam}. The value slope parameter ($L_0 = 87$~MeV) is also within the range of $L_0$ prescribed by \cite{Dutra:2014qga} and comparable with the results of \cite{Fattoyev:2017jql,Zhu:2018ona} based on the co-relations between the symmetry energy and tidal deformability and radius of a 1.4 $M_{\odot}$ NS.
 
  However, as seen from \cite{Jha:2009kt,Jha:2006zi,Jha:2007ej} that the EoS yielded by this parameter set, though is in accordance with the heavy-ion collision data \cite{Danielewicz:2002pu}, passes through the soft band of heavy-ion collision data for PNM \cite{Jha:2009kt}. As discussed in \cite{Sahu:2000ut,Jha:2009kt,Sen:2018yyq}, this is because of the high value of nucleon effective mass ($m^*=0.87~ m$) yielded by the present model compared to other RMF models \cite{Horowitz:1981xw,Dutra:2014qga} and the dominance of vector repulsive force at high densities. It can be seen from table 1 of \cite{Jha:2009kt} that higher ratio of $C_{\sigma}/C_{\omega}$ and more negative values of $B$ give higher effective mass and lower nuclear incompressibility and consequently softer EoS \cite{Jha:2009kt} yielding low mass NS configurations that do not satisfy the maximum mass constraint of NSs \cite{Cromartie:2019kug}. 
 
 Next we present the parameter sets chosen for the dark matter sector in section \ref{Param_DM}.

\subsection{Parameter set for dark sector}
\label{Param_DM}

 Inside the NS we are not considering non-thermal production of DM fermions ($\chi, \bar{\chi}$). To set some benchmark points, we consider the self interaction constraints coming from bullet cluster to determine the corresponding value of the mass ($m_\phi$) of the light mediator $\phi$~\cite{Tulin:2013teo, Tulin:2017ara,Hambye:2019tjt}. To determine the value of the coupling $y_\phi$ we took the help of the present day thermal relic abundances of DM~\cite{Belanger:2013oya, Gondolo:1990dk, Guha:2018mli}. So in case of non-thermal production this is not stringent.  

\begin{table}
\caption{Chosen values of self interacting DM $m_\chi$  and corresponding values of $m_\phi$ from the constraints obtained from Bullet cluster.  $y_\phi$ has been fixed from observed relic abundance.}
{{
\setlength{\tabcolsep}{25.5pt}
\begin{center}
\begin{tabular}{ c c c c c c c c}
\hline
\hline
 $m_{\chi}$ & $m_{\phi}$ & $y_{\phi}$ \\
 (GeV) &(MeV)  &  \\

\hline
\hline
$1$ & $4$ & $0.06$ \\
$5$ & $9$ & $0.11$ \\
$10$ & $15$ & $0.15$ \\  
$15$ & $20$ & $0.18$ \\  
$20$ & $30$ & $0.2$ \\  
$50$ & $60$ & $0.32$ \\  
\hline\hline
\end{tabular}
\end{center}
}}
\protect\label{table_DM}
\end{table}
 
\begin{figure}
\centering
\includegraphics[width=0.5\textwidth]{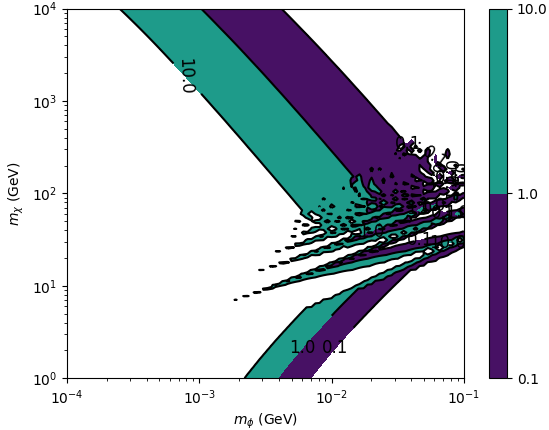}
\caption{Combination of $m_\chi$ and $m_\phi$ satisfying the self-interaction constraint from bullet cluster~\protect\cite{Randall:2007ph, Tulin:2013teo}. The color coding denotes the values of DM self-interaction $\sigma_T/m_\chi = (0.1\rm{-}1)~\rm{cm^2/gm}$(violet) and $(1\rm{-}10)~\rm{cm^2/gm}$(green).}
\label{bullet_cluster}
\end{figure}

 For DM fermions of mass $m_\chi$, self-scattered through a light scalar mediator of mass $m_\phi$, the constraint from bullet cluster structure formation is $\sigma_T/m_\chi \leq 1.25~\rm{cm^2/gm}$~\cite{Randall:2007ph, Robertson:2016xjh}; $\sigma_T$ be the self-scattering transfer cross-section. In figure \ref{bullet_cluster} we show the region $\sigma_T/m_\chi = (0.1\rm{-}10)~\rm{cm^2/gm}$, which is the typical value for known galaxies and clusters~\cite{Randall:2007ph,Bradac:2006er,Dawson:2011kf,Dave:2000ar,Vogelsberger:2012ku,Kahlhoefer:2015vua}. The coupling $y_\phi$ has been fixed from observed relic abundance~\cite{Belanger:2013oya}. We take $\lambda_{\phi}=0$ in the present work.

 For the analysis we have assumed a constant density of fermionic DM throughout the NS and beyond. In our analysis, the value of the DM density is $\sim 1000$ times smaller than  the average neutron number density. DM mass fraction inside the NS is $\sim \mathcal{O}(0.01)$~\cite{Li:2012qf, Panotopoulos:2017idn}. Using the SNM number density, the DM number density becomes roughly $\rho_{\chi} = 10^{-3}\times \rho_0 \sim 0.15\times 10^{-3}~\rm{fm^{-3}}$. From that we estimated a constant Fermi momentum of dark matter fermions to be $k_F^{\chi} = 0.033~\rm{GeV}$. For $10\rho_0$, $k_F^{\chi}$ is around $0.06~\rm{GeV}$. So we varied $k_F^{\chi}$ from $0.01~\rm{GeV}$ to $0.06~\rm{GeV}$.
 
 Considering the aforesaid parameter sets for both hadronic matter and DM, we proceed to compute the EoS and structural properties of NS in presence of DM in the next section. The structural properties of NS are obtained by solving the Tolman-Oppenheimer-Volkoff (TOV) equations \cite{Tolman:1939jz, Oppenheimer:1939ne} numerically with the obtained EoS. The dimensionless tidal deformability ($\Lambda$) can be obtained in terms of the mass, radius and the tidal love number ($k_2$) following \cite{Hinderer:2007mb,Hinderer:2009ca,Alvarez-Castillo:2018pve}. The dimensionless tidal deformability of the individual components of the BNS associated with merger corresponding to GW170817 can also be calculated following \cite{Raithel:2018ncd,Alvarez-Castillo:2018pve}.

\section{Results and Discussions}
\label{Results}

\subsection{Dark matter admixed neutron star for different values of $m_{\chi}$} 
\label{EOSmchi} 

\begin{figure}
\centering
\includegraphics[width=0.5\textwidth]{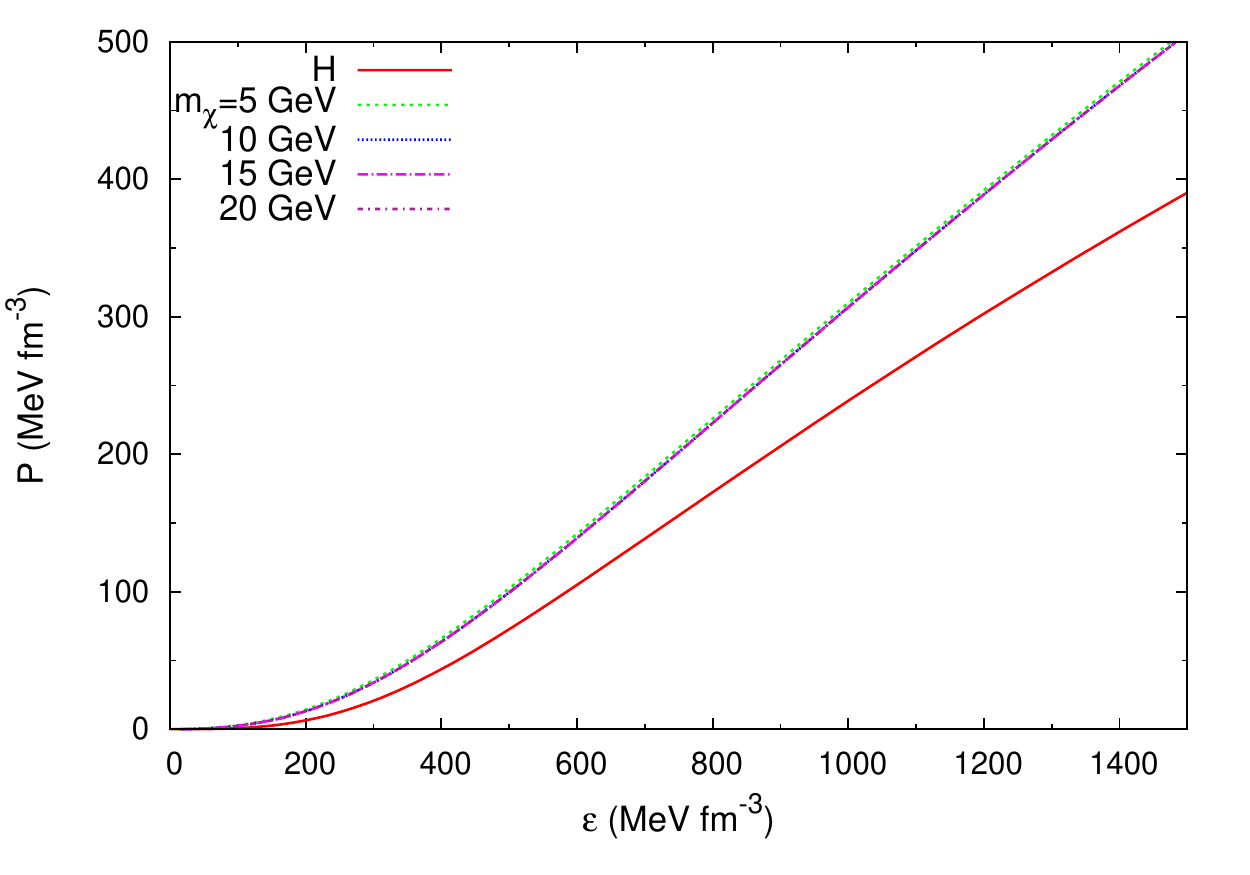}
\caption{Equation of state of neutron star with $\beta$ stable matter without dark matter (H) and dark matter admixed neutron star matter for different values of $m_\chi$.}
\label{eos1}
\end{figure}

Firstly, we compute the DM admixed EoS for different values of $m_{\chi}$ for $k_F^{\chi} = 0.06~\rm{GeV}$. The result is depicted in figure \ref{eos1}.

\begin{figure}
\centering
\includegraphics[width=0.5\textwidth]{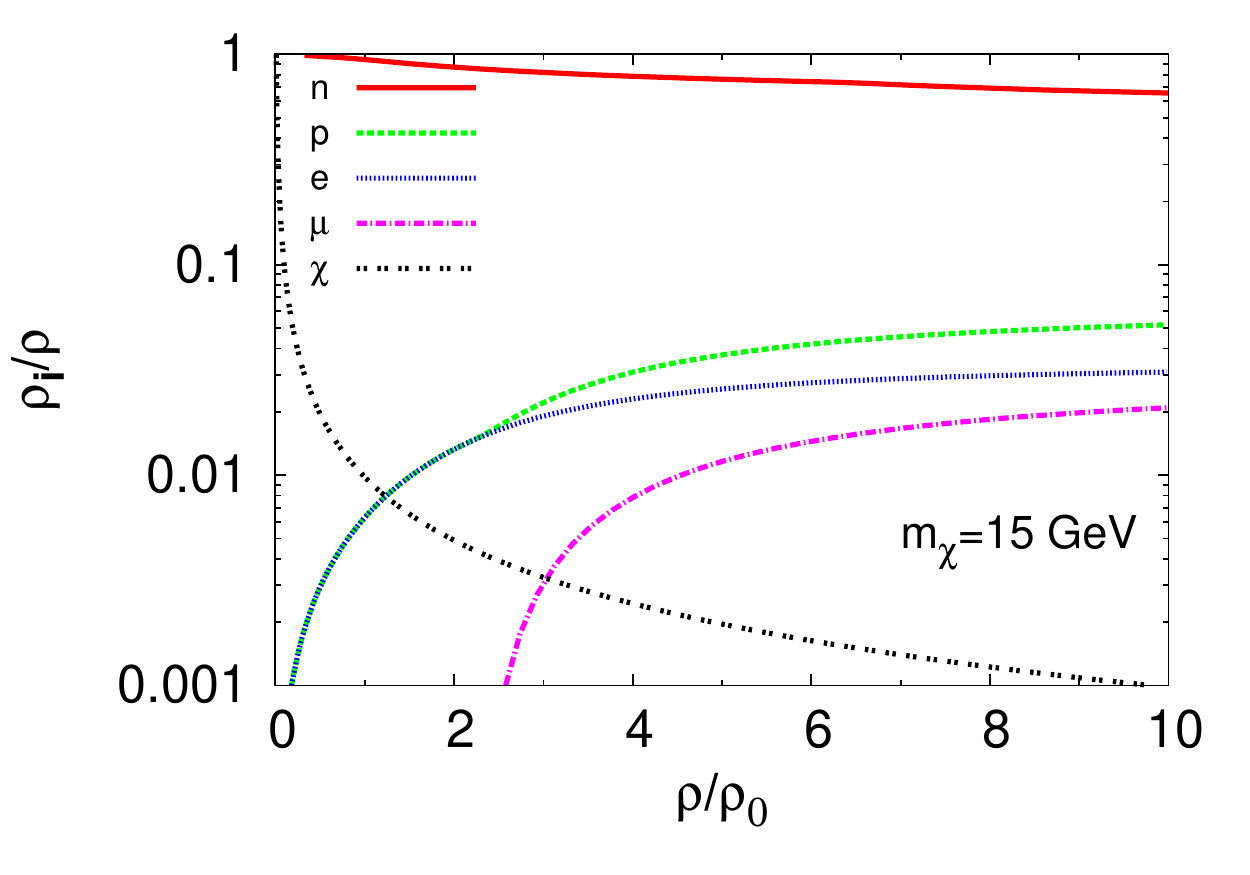}
\caption{Relative particle population of dark matter admixed neutron star matter for $m_\chi$=15 GeV.}
\label{pf}
\end{figure}

 For better comparison, we have also shown the EoS for $\beta$ equilibrated NSM without DM as `H'. It is seen that the EoS stiffens considerably when the presence of DM is considered compared to the pure hadronic EoS. With the decrease in DM mass, further feeble stiffening is noticed. Contrary to the case where massive (200 GeV) DM is chosen as a possible constituent of NS and consequently the EoS softens compared to the no-DM case \cite{Panotopoulos:2017idn}, we find that comparatively lighter and feeble interaction of DM with the nucleons stiffens the EoS considerably. This stiffening of EoS in presence of DM can be attributed to the fact that light and the feeble interaction between the DM particles with the nucleons do not affect the nucleon momenta and particle population much and therefore the EoS for the hadronic sector is very less affected while the total energy density and pressure increase due to DM contribution to them as seen in eqs. \ref{e} and \ref{P}. To understand this fact better we show in figure \ref{pf} the relative abundance of different particles in the NSM including DM.

 The neutron fraction reduces as protons, electrons and muon concentration increase, guided by the chemical equilibrium and charge neutrality conditions for $\beta$ equilibrium. The DM fraction reduces gradually towards the core and the hadronic matter population remains almost unaffected by the presence of DM swallowed during the formation of NS. This happens due to the consideration that DM has a constant number density throughout the NS. The correspondence is justified by the presence of the dense DM halo as one move outwards NS \cite{Li:2012ii}. Inside the core, at a particular density, moderate fraction of DM adds to the net pressure without affecting the contribution from the hadronic sector and therefore we observe stiffening of the EoS in presence of DM. The density of the DM halo is dependent on the environment at which the NS is formed and is predicted by N-body simulations via NFW profile of various galaxies and clusters~\cite{Navarro:1995iw}.

\begin{figure}
\centering
\includegraphics[width=0.5\textwidth]{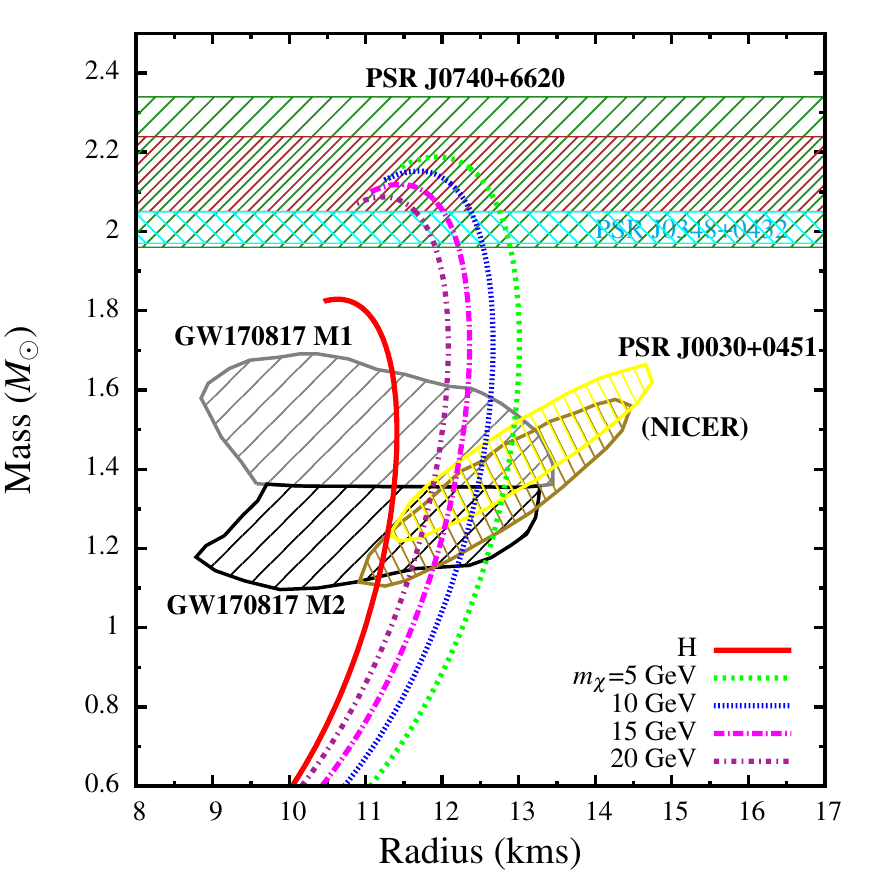}
\caption{Mass-radius relationship of static neutron stars with $\beta$ stable matter without dark matter (H) and dark matter admixed neutron star matter for different values of $m_\chi$. Observational limits imposed from high mass pulsars like PSR J0348+0432 ($M = 2.01 \pm 0.04~ M_{\odot}$) \protect\cite{Antoniadis:2013pzd} (cyan shaded region) and PSR J0740+6620 ($2.14^{+0.10}_{-0.09}~ M_{\odot}$ (68.3\% - brown shaded region) and $2.14^{+0.20}_{-0.18}~ M_{\odot}$ (95.4\% - dark green shaded region)) \protect\cite{Cromartie:2019kug} are also indicated. The limit on $R_{1.4}$ \protect\cite{TheLIGOScientific:2017qsa,Abbott:2018exr,Fattoyev:2017jql}  prescribed from GW170817 are indicated by the black horizontal line with arrows. The constraints on $M-R$ plane from NICER experiment for PSR J0030+0451 are also compared (golden shaded region \protect\cite{Riley:2019yda} and yellow shaded region \protect\cite{Miller:2019cac}.}
\label{mr1}
\end{figure}

 The structural properties like gravitational mass, radius and surface redshift of DM admixed NS are computed next for the obtained EoS for different values of $m_{\chi}$. In figure \ref{mr1} we present the variation of gravitational mass $M$ with respect to radius $R$. We have also compared the case when pure hadronic matter is considered without DM (`H').

 As the EoS stiffens with the inclusion of DM, the effect is also evident from the mass-radius relationship. The maximum gravitational mass has increased considerably when the contribution of DM is considered for any given value of $m_\chi$ compared to the pure hadronic case (`H'). It is also seen that lighter DM fermions contribute more to both mass and radius of NS. For every value of $m_\chi$ considered, the maximum mass constraints from the massive pulsars like PSR J0348+0432 \cite{Antoniadis:2013pzd} and PSR J0740+6620 \cite{Cromartie:2019kug} are satisfied with the DM admixed NS EoS. It is noteworthy that the pure hadronic EoS alone in absence of DM could not satisfy the maximum mass constraints from these two massive pulsars. Thus the consideration of DM contribution in NS could successfully yield NS configurations compatible with massive pulsar observations. This implies that the model parameter set (shown in table \ref{table1}), that was ruled out (despite satisfying the SNM properties reasonably well) because it could not yield high mass NS configurations, can be revived by considering the presence of light and feebly interacting DM as a possible constituent of NS. Moreover, we also found that the NICER data for mass and radius of PSR J0030+0451 \cite{Riley:2019yda,Miller:2019cac} are better satisfied with all the values of $m_\chi$ considered in the present work compared to the pure hadronic matter case (figure \ref{mr1}). The bound on $R_{1.4}$ from GW170817 \cite{TheLIGOScientific:2017qsa,Abbott:2018exr,Fattoyev:2017jql} is well satisfied for all the cases (with and without DM). In this work we are interested to show that appropriately small values of fermionic DM mass can help to obtain reasonable NS configurations. However, we have also shown that the higher values of $m_{\chi}$ will led to gradual softening of EoS and consequently reduction in maximum mass, radius and tidal deformability of NSs (table \ref{table1}). Therefore considering $m_\chi$ as high as 200 GeV will inevitably soften the EoS and low mass NS compared to the no-DM case as shown in \cite{Panotopoulos:2017idn} and other works.

\begin{figure}
\centering
\includegraphics[width=0.5\textwidth]{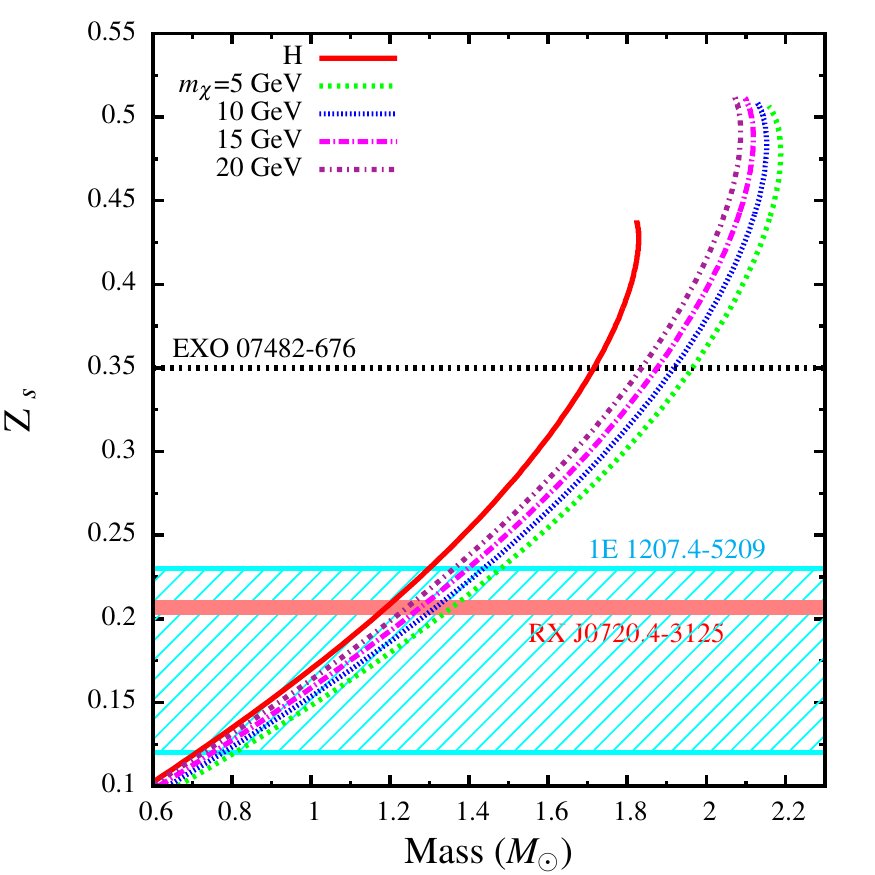}
\caption{Surface gravitational redshift $Z_s$ vs gravitational mass $M$ of dark matter admixed neutron star for different values of $m_\chi$. Observational limits imposed from EXO 07482-676 ($Z_S = 0.35$) \protect\cite{Cottam:2002cu}, 1E 1207.4-5209 ($Z_S = (0.12 - 0.23)$) \protect\cite{Sanwal:2002jr} and RX J0720.4-3125 ($Z_S = 0.205_{-0.003}^{+0.006}$ \protect\cite{Hambaryan:2017wvm} are also indicated.}
\label{mZ}
\end{figure}

 The variation of surface redshift $Z_s$ with respect to gravitational mass for DM admixed NS is depicted in figure \ref{mZ} for the three values of $m_\chi$. 
 
 There is considerable increase in the value of maximum surface redshift when DM is considered compared to the pure hadronic case (`H'). There is feeble increase in the value of maximum $Z_s$ with the variation of $m_\chi$. Although $m_\chi$=100 MeV and 1 GeV yield the same value of maximum mass but the maximum redshift for the latter is slightly greater than that of the former. This is because along with the mass of NS, the surface redshift is also dependent on the value of radius which is slightly less in the latter case i.e., $m_\chi$=1 GeV. The maximum surface redshift constraints from EXO 07482-676 \cite{Cottam:2002cu}, 1E 1207.4-5209 \cite{Sanwal:2002jr} and RX J0720.4-3125 \cite{Hambaryan:2017wvm} are well satisfied with both DM admixed and pure hadronic NS EoS.
  
\begin{figure}
\centering
\includegraphics[width=0.5\textwidth]{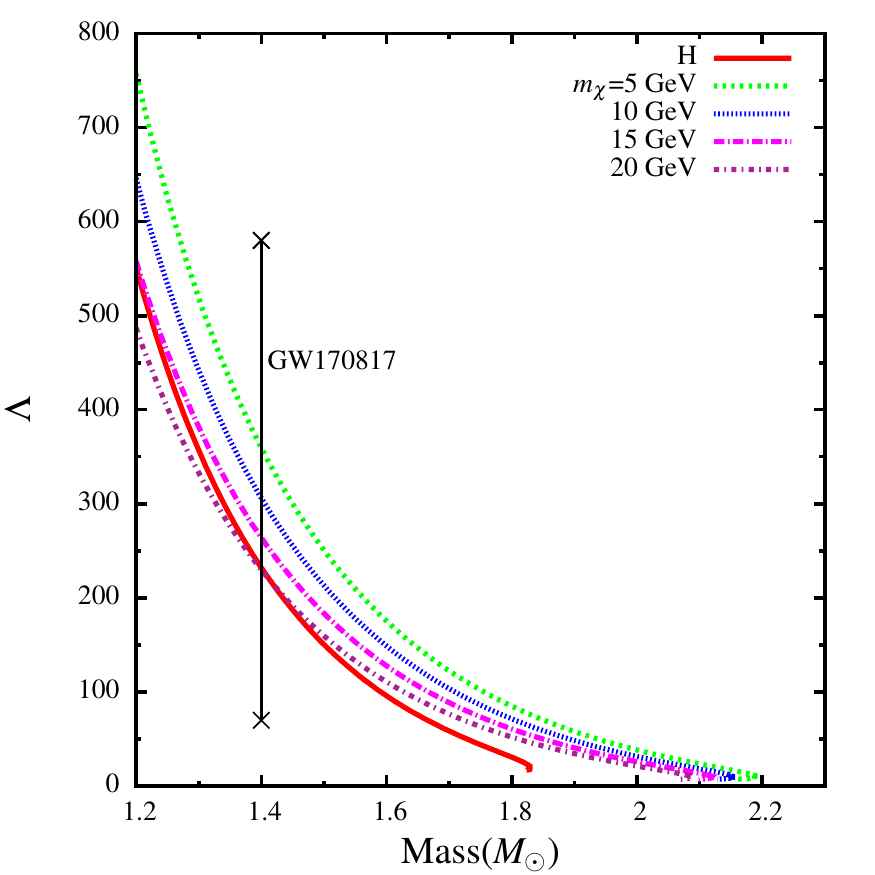}
\caption{Variation of tidal deformability with respect to gravitational mass of neutron stars with $\beta$ stable matter without dark matter (H) and dark matter admixed neutron star matter for different values of $m_\chi$. Constraint on $\Lambda_{1.4}$ from GW170817 observations is also indicated following \protect\cite{TheLIGOScientific:2017qsa,Abbott:2018exr}.}
\label{mLam1}
\end{figure}

 We next calculate the dimensionless tidal deformability $\Lambda = \frac{2}{3} k_2 (\frac{M}{R})^{-5} $ and present our results in figure \ref{mLam1}. For a spherically symmetric star, physical significance of the tidal deformability lies in the modification of the spacetime metric by a linear $l = 2$ perturbation.
 
 As expected, the tidal deformability decreases with increasing mass, indicating that massive stars are less deformed. There are decrements in the values of $\Lambda$ for  increasing values of $m_\chi$. This is because the compactness ($C\equiv M/R$), that increases with $m_\chi$, affects $\Lambda$ directly. Our estimates of $\Lambda_{1.4}$ are in excellent agreement with the constraint on the same obtained from GW170817 data analysis~\cite{TheLIGOScientific:2017qsa,Abbott:2018exr}.
 
\begin{figure}
\centering
\includegraphics[width=0.5\textwidth]{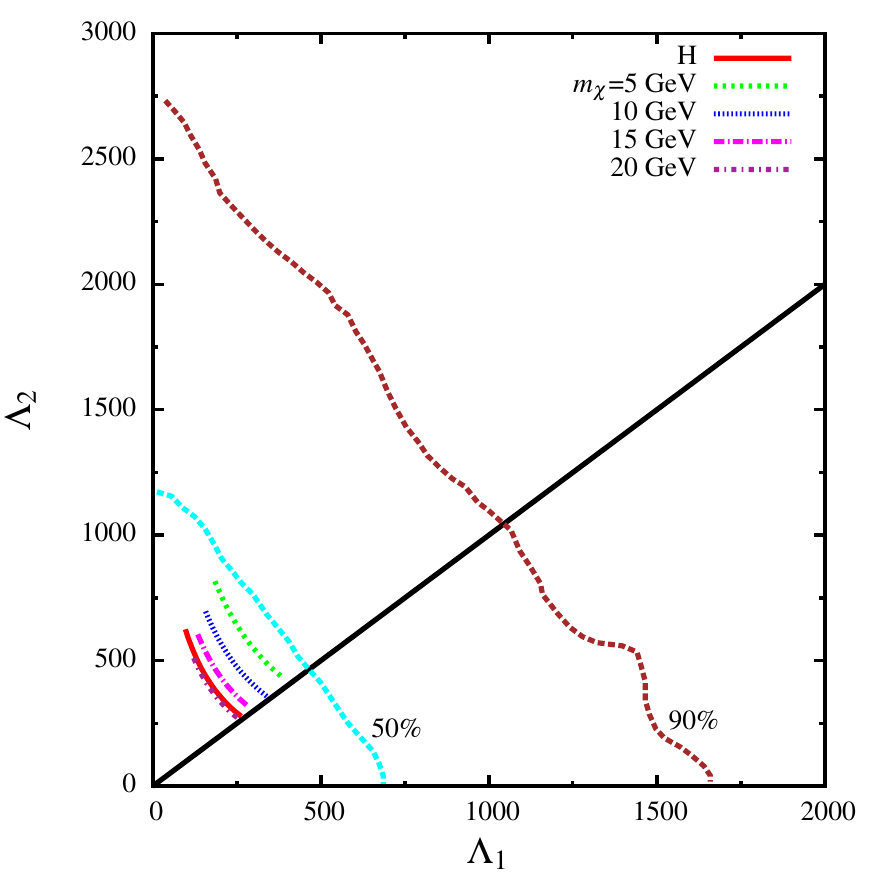}
\caption{Tidal deformabilities of the individual components of the binary neutron stars associated with GW170817 with $\beta$ stable matter without dark matter (H) and dark matter admixed neutron star matter for different values of $m_\chi$. The 50\% and 90\% confidence limits for this event are also indicated following \protect\cite{TheLIGOScientific:2017qsa,Abbott:2018exr}.}
\label{l1l2}
\end{figure}

 In figure \ref{l1l2} we show the variation of tidal deformability parameters $\Lambda_1$ and $\Lambda_2$ linked to the BNS companion with a high mass $M_1$ and a low mass $M_2$ associated with GW170817 observation. This result is obtained considering NS both with and without DM. The observed mass range for the BNS companion is $1.1 M_\odot \leqslant M_{NS} \leqslant 1.6 M_\odot $. To ensure the fact $M_2 < M_1$ for the figure \ref{l1l2}, we used $1.365 M_\odot \leqslant M_{1} \leqslant 1.6 M_\odot $ and fixed $M_2$ from the observed chirp mass $M_{chirp} = 1.188 M_\odot$  from GW170817 data.

 Our estimates of $\Lambda_1$ and $\Lambda_2$ both in presence and absence of DM are within the bounds specified from GW170817 data analysis \cite{TheLIGOScientific:2017qsa,Abbott:2018exr}.
 
 The various structural properties of NS obtained with and without considering DM are tabulated below in table \ref{table1}.

\begin{table*}
\begin{center}
\caption{Structural properties of static neutron stars with $\beta$ stable matter without dark matter (H) and dark matter admixed neutron star matter (H+DM) for different values of $m_\chi$.}
\setlength{\tabcolsep}{15.0pt}
\begin{center}
\begin{tabular}{cccccccc}
\hline
\hline
\multicolumn{1}{c}{} &
\multicolumn{1}{c}{$M_{max}$}  &
\multicolumn{1}{c}{$R$} &
\multicolumn{1}{c}{$R_{1.4}$} &
\multicolumn{1}{c}{$\Lambda_{1.4}$} &
\multicolumn{1}{c}{$Z_s$}\\
\multicolumn{1}{c}{} &
\multicolumn{1}{c}{($M_{\odot}$)} &
\multicolumn{1}{c}{(km)} &
\multicolumn{1}{c}{(km)} & 
\multicolumn{1}{c}{} & 
\multicolumn{1}{c}{} \\
\hline
H  &1.83  &10.62  &11.39  &233.50  &0.438 \\
\hline
\hline
\multicolumn{1}{c}{} &
\multicolumn{1}{c}{$m_\chi$} &
\multicolumn{1}{c}{$M_{max}$} &
\multicolumn{1}{c}{$R$} &
\multicolumn{1}{c}{$R_{1.4}$} &
\multicolumn{1}{c}{$\Lambda_{1.4}$} &
\multicolumn{1}{c}{$Z_s$}\\
\multicolumn{1}{c}{} &
\multicolumn{1}{c}{(GeV)} &
\multicolumn{1}{c}{($M_{\odot}$)} &
\multicolumn{1}{c}{(km)} &
\multicolumn{1}{c}{(km)} & 
\multicolumn{1}{c}{} &
\multicolumn{1}{c}{} \\
\hline
H+DM  &0.1  &2.22    &12.20  &13.24   &428.05  &0.503 \\
      &1    &2.22    &12.18  &13.18   &394.33  &0.504 \\
      &5    &2.19    &11.93  &12.86   &346.99  &0.507 \\
      &10   &2.15    &11.67  &12.50   &310.45  &0.509 \\
      &15   &2.12    &11.45  &12.21   &256.92  &0.511 \\
      &20   &2.09    &11.23  &11.92   &231.96  &0.513 \\
      &50   &1.92    &10.18  &10.69   &112.59  &0.518 \\
\hline
\hline
\end{tabular}
\end{center}
\protect\label{table1}
\end{center}
\end{table*}

 The mass of fermionic DM $m_{\chi}$ therefore plays a very significant role in determining the EoS and structural properties of DM admixed NS.

\subsection{Dark matter admixed neutron star for different values of $k_F^{\chi}$} 
\label{EOSkfchi} 

\begin{figure}
\centering
\includegraphics[width=0.5\textwidth]{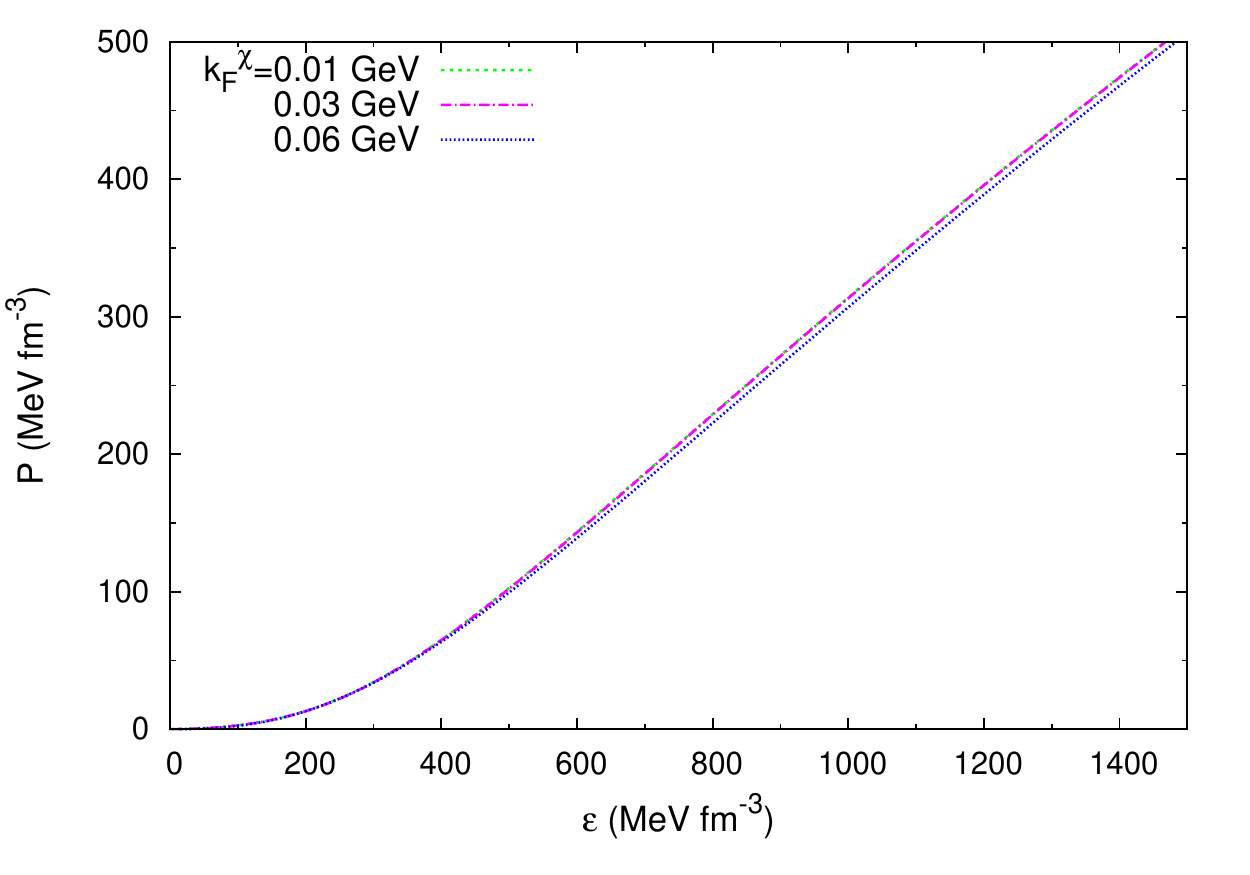}
\caption{Equation of state of dark matter admixed neutron star matter for different values of $k_F^{\chi}$.}
\label{eos2}
\end{figure}

 In order to study the explicit dependence of the EoS and structural properties of NS on the momentum of fermionic DM, we now vary $k_F^{\chi}$ by fixing $m_{\chi}$=10 GeV. The EoS for DM admixed NSM for different values of $k_F^{\chi}$ is shown in figure \ref{eos2}.

\begin{figure}
\centering
\includegraphics[width=0.5\textwidth]{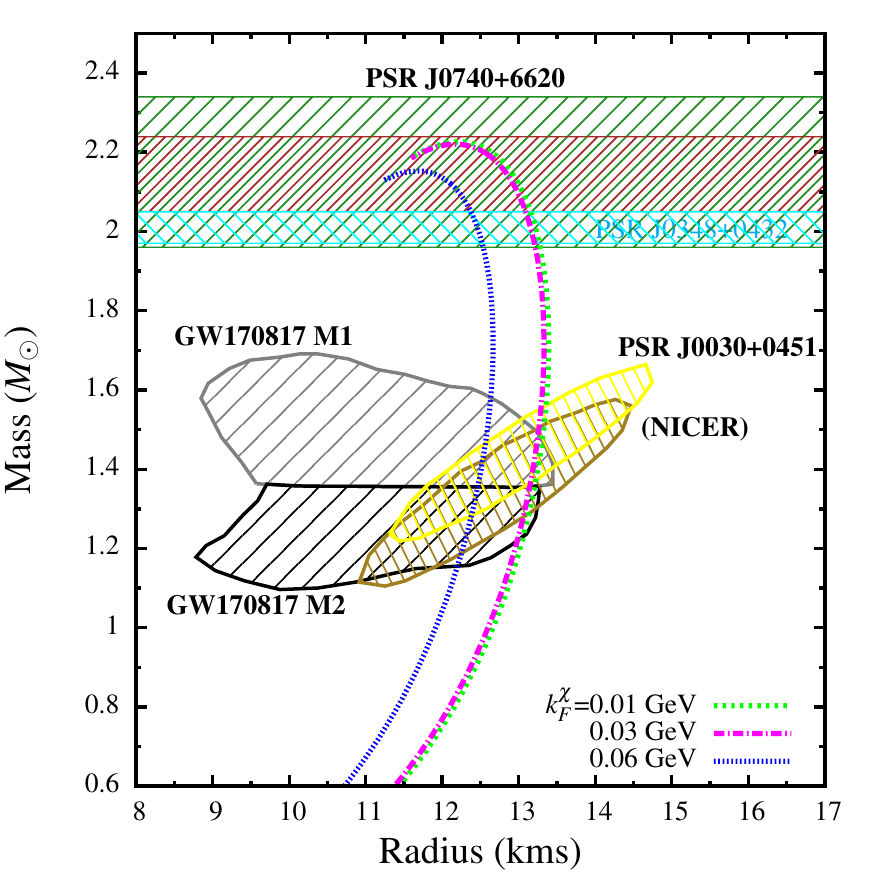}
\caption{Mass-radius relationship of static neutron stars with dark matter admixed neutron star matter for different values of $k_F^{\chi}$. The various constraints shown are same as figure \ref{mr1}.}
\label{mr2}
\end{figure}

 Consistent with \cite{Panotopoulos:2017idn} we find that the EoS stiffens with decreasing values of $k_F^{\chi}$ which directly affects the EoS as seen from eqs. \ref{e} and \ref{P}. The decrease is, however, less in the present work as we have chosen the value of $m_{\chi}$ by satisfying the self-interaction constraint from bullet cluster \cite{Randall:2007ph, Tulin:2013teo} (as shown in figure \ref{bullet_cluster}) and to be quite small compared to that chosen in \cite{Panotopoulos:2017idn}. With the obtained EoS, we also calculated the structural properties of DM admixed NS for varying $k_F^{\chi}$. Figure \ref{mr2} shows the variation of gravitational mass with radius for different values of $k_F^{\chi}$.

 Both radius and mass increase with decreasing value of $k_F^{\chi}$. The maximum gravitational mass constraints from both PSR J0348+0432 \cite{Antoniadis:2013pzd} and PSR J0740+6620 \cite{Cromartie:2019kug} are satisfied with all the chosen values of $k_F^{\chi}$ while the $M-R$ results for all the values of $k_F^{\chi}$ are in excellent agreement with the NICER data for PSR J0030+0451 \cite{Riley:2019yda,Miller:2019cac}. Also the estimates of $R_{1.4}$ are in agreement with the GW170817 data for BNSM \cite{TheLIGOScientific:2017qsa,Abbott:2018exr,Fattoyev:2017jql}.

\begin{figure}
\centering
\includegraphics[width=0.5\textwidth]{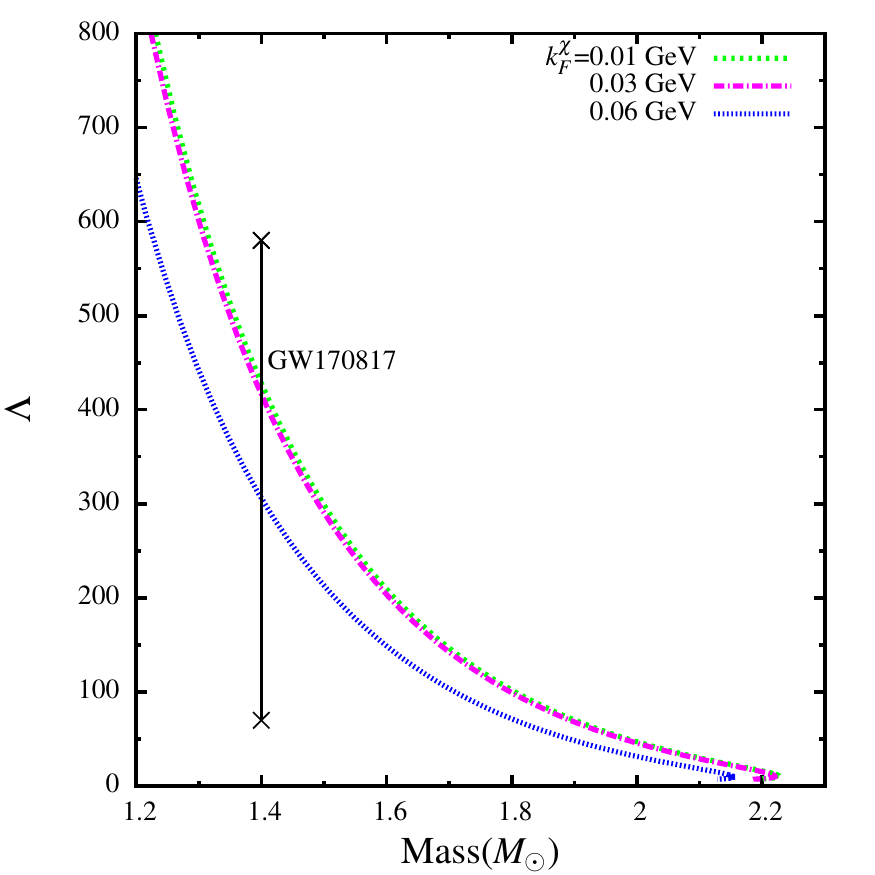}
\caption{Variation of tidal deformability with respect to gravitational mass of neutron stars with dark matter admixed neutron star matter for different values of $k_F^{\chi}$. Constraint on $\Lambda_{1.4}$ from GW170817 observations is also indicated following \protect\cite{TheLIGOScientific:2017qsa,Abbott:2018exr}.}
\label{mLam2}
\end{figure}

 In figure \ref{mLam2} we show the variation of $\Lambda$ with respect to mass 
for different values of $k_F^{\chi}$.

 We found that as the EoS stiffens with decreasing values of $k_F^{\chi}$, $\Lambda$ increases and the value of $\Lambda_{1.4}$ for all the considered values of $k_F^{\chi}$ are consistent with the bound obtained from GW170817 data analysis~\cite{TheLIGOScientific:2017qsa,Abbott:2018exr}.

 Below we tabulate in table \ref{table2} the estimates of various structural properties for different values of $k_F^{\chi}$.

\begin{table}
\begin{center}
\caption{Structural properties of static neutron stars with dark matter admixed neutron star matter for different values of $k_F^{\chi}$.}
\setlength{\tabcolsep}{15.0pt}
\begin{center}
\begin{tabular}{cccccccc}
\hline
\hline
\multicolumn{1}{c}{$k_F^{\chi}$} &
\multicolumn{1}{c}{$M_{max}$} &
\multicolumn{1}{c}{$R$} &
\multicolumn{1}{c}{$R_{1.4}$} &
\multicolumn{1}{c}{$\Lambda_{1.4}$} \\
\multicolumn{1}{c}{(GeV)} &
\multicolumn{1}{c}{($M_{\odot}$)} &
\multicolumn{1}{c}{(km)} &
\multicolumn{1}{c}{(km)} &
\multicolumn{1}{c}{} \\
\hline
0.01   &2.23    &12.23  &13.25  &427.34 \\
0.03   &2.22    &12.17  &13.18  &393.12 \\
0.06   &2.15    &11.67  &12.50  &310.45 \\
\hline
\hline
\end{tabular}
\end{center}
\protect\label{table2}
\end{center}
\end{table}

\hspace{1cm}

\subsection{Enclosed Mass vs Radius Profiles for different values of $k_F^{\chi}$ and $m_{\chi}$}

\begin{figure*}
\centering
\includegraphics[width=0.33\textwidth]{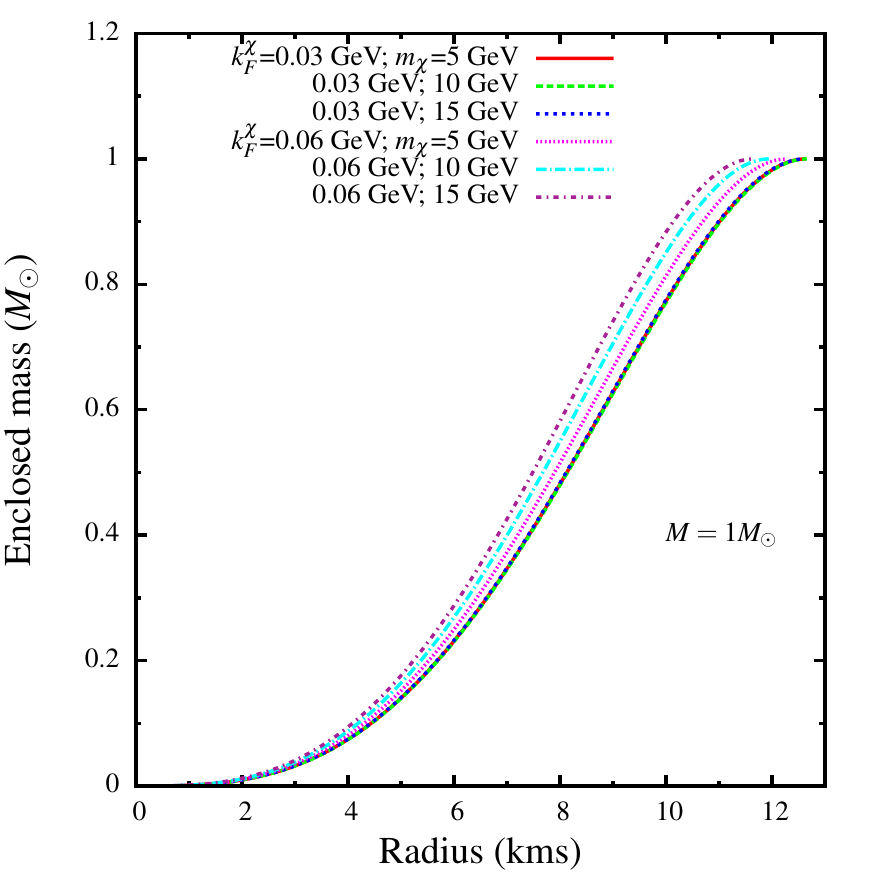}\hfill
\includegraphics[width=0.33\textwidth]{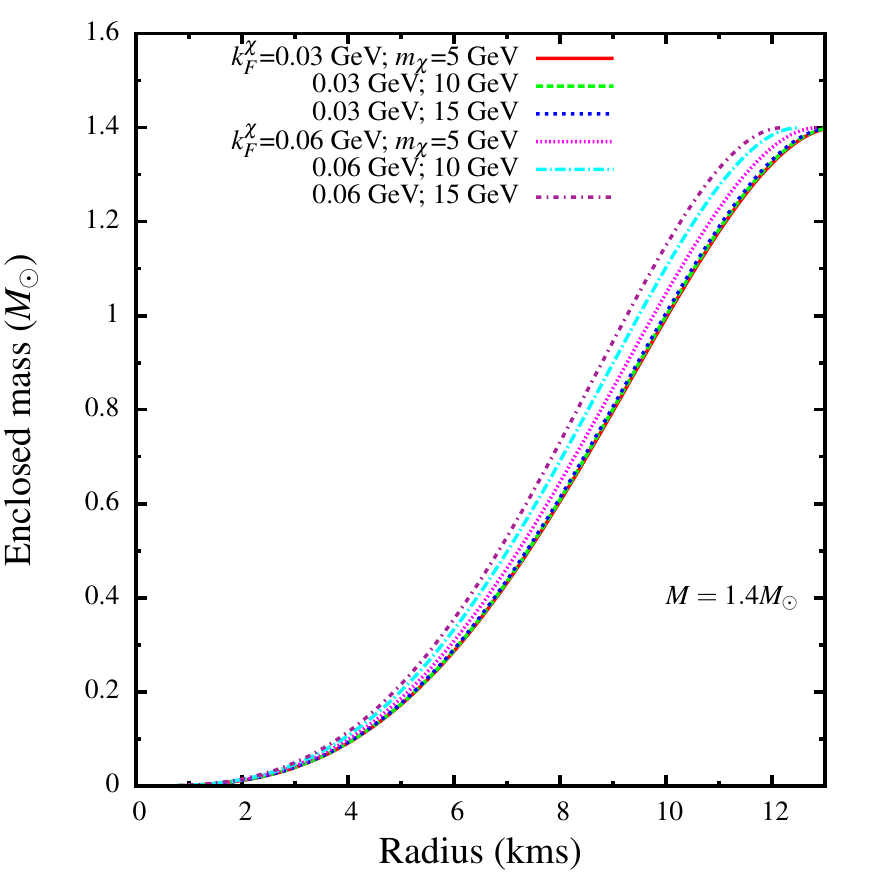}\hfill
\includegraphics[width=0.33\textwidth]{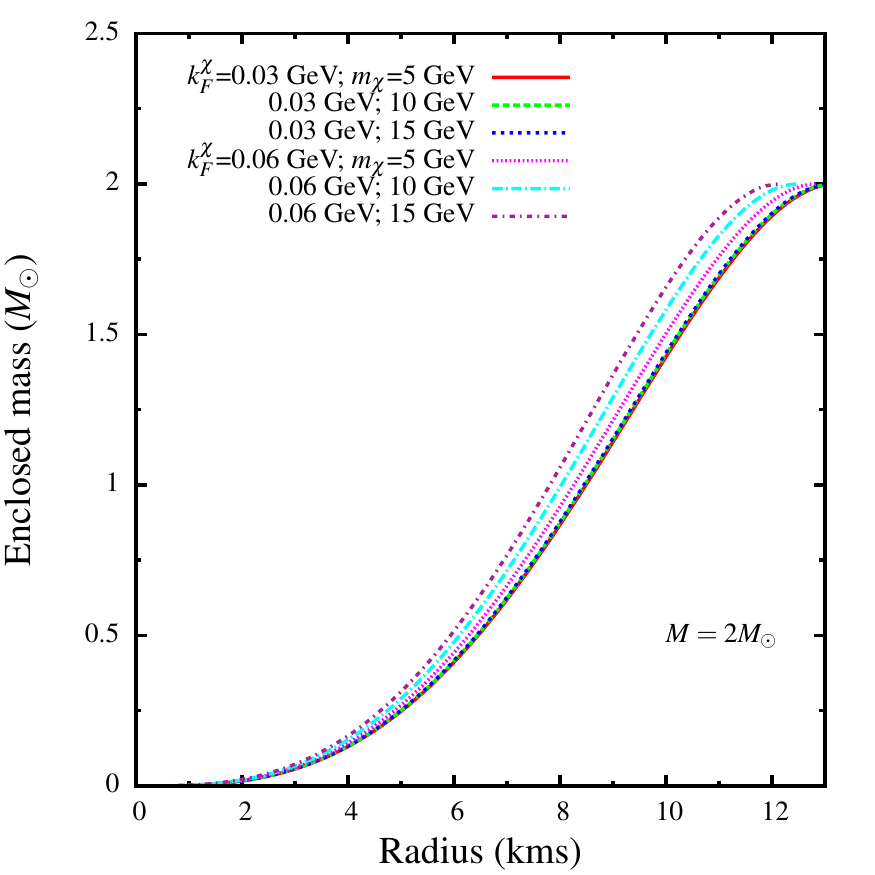}
\caption{Enclosed Mass-radius profile of static neutron stars with dark matter admixed neutron star matter for different values of $k_F^{\chi}$ and $m_{\chi}$. Mass of the NS has been fixed to $M = 1.0 M_\odot , 1.4 M_\odot , 2.0 M_\odot$, respectively.}
\label{mrprofile}
\end{figure*}

 The enclosed mass-radius profile for a chosen value of total NS mass \cite{Molla:2020rnu,DelPopolo:2020hel,Bhat:2019tnz} can be obtained by solving the TOV equations  \cite{Tolman:1939jz, Oppenheimer:1939ne} following \cite{Molla:2020rnu}. In the figure \ref{mrprofile}, we show the enclosed mass profiles as a function of radius for NS mass $M = 1.0 M_\odot , 1.4 M_\odot , 2.0 M_\odot$ for different values of $k_F^{\chi}$ and $m_{\chi}$. 

\section{Summary and Conclusions}
\label{Conclusions}
 
 We consider feebly interacting, thermal DM as a possible constituent of $\beta$ equilibrated NSM. The DM-SM interaction is invoked though light new physics mediator whose mass $m_{\phi}$ and coupling $y_{\phi}$ are relate to the mass of fermionic DM $m_{\chi}$ via the self-interaction constraint from the Bullet cluster and from the present day relic abundance. We show that a suitable choice of low GeV $m_{\chi}$, consistent with the constraints from Bullet cluster, can stiffen the overall EoS. With such consideration, even a soft and ruled out pure hadronic EoS could be successfully revived to obtain static NS structural properties that satisfy the various constraints on NS properties. For all the considered values of $m_{\chi}$, the maximum mass obtained is consistent with the limits imposed from massive pulsars like PSR J0348+0432 and PSR J0740+6620. Our results of $R_{1.4}$ and $\Lambda_{1.4}$, $\Lambda_1$ and $\Lambda_2$ fall within the individual range prescribed by data analysis of GW170817. Moreover, the consideration of DM also helped to satisfy the mass-radius constraints from NICER experiment. Also our estimates of maximum surface redshift are in accordance with the bounds obtained from EXO 07482-676, 1E 1207.4-5209 and RX J0720.4-3125. It is seen that the presence of massive DM fermion reduces the maximum mass, radius and tidal deformability of the NS.

 Since we considered constant density distribution of DM, we also varied the DM Fermi momentum to ensure that for a given low GeV mass value of DM fermion, we obtain stiffer EoS with different values of $k_F^{\chi}$ compared to the no-DM case and consequently reasonable values the maximum mass, radius and tidal deformability compatible with bounds obtained on them from the massive pulsars and GW170817 data.
 
\appendix

\section*{APPENDIX A: Estimation of $\MakeLowercase{g}_{\phi NN}$}

 For the new scalar mediator $\phi$-quark interaction
\bea
\mathcal{L}_{\phi, q} = \sum_q g_q \bar{q} q \phi = \left(\sqrt{2} G_F \right)^{1/2} \sum_q \epsilon_q m_q \bar{q} q \phi
\eea
Here, $G_F$ is the Fermi constant and the parameter $\epsilon_q = 1$ for all quarks in SM but we can vary $\epsilon_q$ for BSM physics~\cite{Cheng:2012qr}.

For nucleons the effective interaction $ g_{\phi N N}  \bar{N} N \phi$ with 
\bea
g_{\phi N N} = \left(\sqrt{2} G_F \right)^{1/2} \sum_q <{N}| \epsilon_q m_q \bar{q} q |{N}>
\eea
 Now 
 \bea
 <{N}| \bar{q} q |{N}> = \frac{m_N}{m_q} f_N^{Sq} ~~~ \rm{for~u,d,s} \nonumber \\
 <{N|} \bar{q} q |{N}> = \frac{2}{27}\frac{m_N}{m_q} f_N^{SG} ~~~ \rm{for~c,b,t}
 \eea
 We can estimate $g_{\phi N N}$ with one free parameter $\epsilon_q$, whereas, we have the following standard values from lattice QCD simulation data and chiral perturbation theory~\cite{Backovic:2015cra}
\bea
f_p^{Su} &=& 0.0153; ~~f_p^{Sd} = 0.0191; ~~ f_p^{Ss} = 0.0447; \nonumber \\
f_n^{Su} &=& 0.0110; ~~f_n^{Sd} = 0.0273; ~~ f_n^{Ss} = 0.0447; \nonumber \\
f_N^{SG} &=& 1 - \sum_q f_N^{Sq}
\eea

\textbf{Estimated value} of $g_{\phi NN}$ to be
\bea
g_{\phi NN} \leqslant 1.1 \times 10^{-4}
\eea

\section*{Data availability}

 The data underlying this article are available within the article.

\end{document}